\documentclass[apj]{emulateapj}

\slugcomment{\it Accepted by ApJ --- February 11, 2008}
\shorttitle{Milky Way Rotation Curve and Vertical Derivatives}
\shortauthors{Levine, Heiles, \& Blitz}

\begin{document}

\title{The Milky Way Rotation Curve and its Vertical Derivatives\\ Inside the Solar Circle}
\author{E.S. Levine, Carl Heiles, and Leo Blitz
}
\affil{Department of Astronomy, University of California at Berkeley, Berkeley, CA 94720-3411 USA}
\email{elevine@astron.berkeley.edu}

\begin{abstract}
We measure the Galactic rotation curve and its first two vertical derivatives in the first and fourth quadrants of the Milky Way using the 21 cm VGPS and SGPS. We find tangent velocities of the atomic gas as a function of galactic longitude and latitude by fitting an analytic line profile to the edges of the velocity profiles. The shape of the analytic  profile depends only on the tangent velocity and the velocity dispersion of the gas. 
We use two complementary methods to analyze the tangent velocities: a global model to fit typical parameter values and a local fitting routine to examine spatial variations. We confirm the validity of our fitting routines by testing simple models. Both the global and local fits are consistent with a vertical falloff in the rotation curve of $-$22$\pm6$ km s$^{-1}$ kpc$^{-1}$ within 100 pc of the Galactic midplane.
The magnitude of the falloff is several times larger than what would be expected from the change in the potential alone, indicating some other physical process is important. The falloff we measure is consistent in magnitude with that measured in the halo gas of other galaxies.
\end{abstract}

\keywords{ISM: general --- Galaxy: disk --- Galaxy: fundamental parameters --- Galaxy: kinematics and dynamics --- Galaxy: structure --- radio lines: general}  

\section{Introduction}

The shape of the Milky Way rotation curve is indicative not only of the integrated mass within a radius, but also of the local mass distribution and physical environment. The change in the rotation curve with distance from the plane can tell us about the gravitational potential of the disk as well as which physical processes affect the gas kinematics. Previous studies of H {\footnotesize I} near the plane assumed that the rotation velocity was constant up to $\sim1$ kpc \citep{CRB1979,L1984}; in this paper we test that assumption. 

In other galaxies, the falloff of the halo gas rotation curve with distance from the plane  is well established, having been observed in H {\footnotesize I} (NGC 891: \citealt{SSV1997} and \citealt{OFS2007}, NGC 2403: \citealt{FVSO2002}, NGC 5775: \citealt{LIDC2001}), H$\alpha$ (NGC 891: \citealt{KPDV2007}, NGC 5775: \citealt{R2000a} and \citealt{TDSUR2000}), and X-rays \citep{S2004a}. A recent study of the galactic halos of NGC 891, 4302, and 5775  using H $\alpha$ found that the decrease in the rotation curve with scale height was a nearly constant $-$15 to $-$25 km s$^{-1}$ per scale height across the three galaxies \citep{HRBB2007}. In the Milky Way,  observations of an Ophiuchus superbubble in H {\footnotesize I} and H $\alpha$ at Galactic longitude $\sim30^\circ$ and 3.4 kpc above the plane showed gas that lags 27 km s$^{-1}$ behind the rotation at $b=0$ \citep{PLS2006}. Halo gas motions are likely to be dominated by complex physics, such as a  Galactic fountain \citep{SF1976,B1980} where gas is pushed off of the disk, moves through the halo, and then returns to the disk. \citet{B2002} summarized several different mechanisms of establishing a vertical falloff in the rotation speed, including radial pressure gradients and magnetic tension.

Nearer to the plane in the Milky Way H {\footnotesize I} disk, fluctuations in the velocity structure associated with spiral arms were first observed by \citet{O1962}. These were termed ``rolling motions'' because on one side of the plane there is a positive shift in the tangent velocity, and on the other side the shift is towards more negative velocities. \citet{YW1973} tried to explain the rolling motions as purely a geometric effect, but later work showed that this could not be true in all cases \citep{SP1976,FS1985}.  Rolling motions have also been observed in other spiral arm tracers, such as CO \citep{W1981} and stars \citep{BS1984}.
Note that it is difficult to detect similar phenomena in the disks of external galaxies because $\sim10$ pc spatial resolution and $\sim1$ km s$^{-1}$ velocity resolution are necessary. 

In principle, it is simple to extract the rotation structure of the inner Galaxy from the tangent velocities of gas at different longitudes \citep{KMW1954}; for each longitude, just find the velocity where emission has fallen off. However, this task is complicated by several factors that obfuscate the determination of the tangent velocity, including velocity dispersion due to random motion of the gas, bulk flows often associated with spiral density waves \citep{K1962, SB1966,B1972}, 
and noncircular motions, whether due to the Galactic bar \citep{P1975} or some other cause. 
Nevertheless, H {\footnotesize I} spectra have been used to measure the inner Galaxy rotation curve for more than half a century. 
Early studies also uncovered a large-scale asymmetry between inner Galaxy rotation curves on either side of the Galactic center \citep{K1962}, which may be a sign of a triaxial density distribution \citet{BS1991}.
Much attention has been paid to bumps in the rotation curve as a function of radius \citep{SB1966,B1971,AMB1990,MD2007}, but comparatively little analysis has been done on the fluctuations above and below the plane.

In this paper, we measure the inner Galaxy rotation structure close to the plane using 21 cm emission from H {\footnotesize I} gas.  For the first time, we allow the rotation curve to vary with height off the plane as well as with galactocentric radius.

\section{Formalism}
We  generalize the existing rotation curve formalism to include variation in the rotation speed as a function of height off the plane. 
To accomplish this, we need two coordinate systems. The first is centered at the LSR, with coordinates Galactic longitude $l$, Galactic latitude $b$, and line-of-sight velocity $v_r$.
The second is a cylindrical coordinate system with the origin at the Galactic center, radial coordinate $R$, and height off the plane $z$. 

Analysis of the gas at the tangent points is straightforward because, unlike the rest of the inner Galaxy, distance determinations  in these areas do not suffer from the distance ambiguity.
Furthermore, simple geometric relations can be used to calculate  the galactocentric radius of a tangent point, $R=R_0\left|\sin\ell\right|$, and the height of gas at a tangent point, $z=R_0\cos\ell\tan b$. Since this mapping is one-to-one at a tangent point, we use these two coordinate systems interchangeably. We assume $R_0$, the distance from the Sun to the Galactic center, is 8.5 kpc. 

Assuming circular rotation, the  relation for line-of-sight velocity is given by
\begin{equation}\label{eqn:genrot}
v_{r}=\sin(l)\cos(b)\left[\frac{R_0}{R}\Theta(R,z)-\Theta_0\right],
\end{equation}
where  $\Theta_0$ is the rotation speed of the LSR (which we assume to be 220 km s$^{-1}$), and we have generalized to a rotation curve that varies with height off the Galactic plane,  $\Theta(R,z)$. We assume cylindrical non-intersecting gas orbits, with no velocity in the $z$ direction.  At the tangent points, we then reduce eqn.~(\ref{eqn:genrot}) to an equation for the tangent velocity
\begin{equation}\label{eqn:tan}
v_{t}(l,b)=\cos(b)\left[\pm\Theta(R,z)-\Theta_0\sin(l)\right],
\end{equation}
where the ``$+$'' is for quadrant I (QI, $0^\circ\le l\le90^\circ$) and the ``$-$'' for quadrant IV (QIV, $270^\circ\le l<360^\circ$). We use measurements of $v_t$ to constrain models of $\Theta(R,z)$.

\section{Data}\label{sec:data}

For the longitudes $18^\circ\le l\le67^\circ$, we used the 21cm H {\footnotesize I} VLA Galactic Plane Survey (VGPS) data \citep{VGPS} to find the tangent velocities as a function of $\ell$ and $b$. The VGPS is a combination of interferometric VLA observations with single-dish observations from the Green Bank telescope. The data have a velocity resolution of 1.56 km s$^{-1}$, an angular resolution of 1', and an rms noise level of 2 K.  The $b$ coverage varies with longitude; near $l\sim 20^\circ$  the data are restricted to $|b|\le1\fdg3$, while around $l\sim60^\circ$  the range is $|b|\le2\fdg3$. These data have a higher angular resolution than that of the Leiden/Argentine/Bonn survey, $0.5^\circ$ \citep{LAB}; the better angular resolution provides the necessary resolution in $z$.

For the southern longitudes $270^\circ\le l\le358^\circ$, we used the 21 cm H {\footnotesize I} Southern Galactic Plane Survey (SGPS) data \citep{SGPS}. The SGPS is a combination of interferometric observations with the Australia Telescope Compact Array with single-dish observations from the Parkes Radio telescope. These data have a velocity resolution of 0.8 km s$^{-1}$, an angular resolution of $\sim$2', and an rms noise level of $\sim$1.6 K. The SGPS is limited to $|b|\le1\fdg5$, but to avoid edge effects associated with the interferometer coverage we further restricted to $|b|\le 1\fdg4$.

To reduce these surveys to a more manageable size and minimize the impact of instrumental noise and small Galactic perturbations, we broke the data into $0\fdg5$ segments in $l$ and, for each segment, we constructed a composite spectrum by taking the median value of the brightness temperature at each velocity and for each $b$.  
This procedure reduced the number of data points in $l$ while preserving the sampling in $v_r$ and $b$.  We use this smaller data set for the remainder of this paper.

Finally, we  restrict to lines-of-sight with $|z| \le 100$ pc, to insure that each longitude has nearly uniform coverage in $z$. We discuss this in more detail in \S \ref{sec:syserr}.
These cuts leave us with 37396 spectra in the north and 22897 spectra in the south within the range  3 kpc $\le R\le8$ kpc.

\section{Tangent Velocity Determination}

\subsection{Curve Fitting}\label{sec:fit}
Various methods have been used to determine $v_t$ for an H {\footnotesize I} spectra. For example, \citet{M1995} chose the velocity where the profile crosses the 10 K threshold, \citet{K1962} chose the velocity of the first maximum in the spectra, \citet{K1964} used the velocity where the brightness temperature crosses half the maximum, and \citet{SB1966} calculated  $v_t$ using the integral of the spectrum past the first maximum. Even the best of these methods requires correction, as modeling shows that the equivalent width velocity or the half max velocity differs from the actual tangent velocity by $\sim5$ km s$^{-1}$ \citep{BG1978}. Also, functions such as Gaussians \citep{KF1985} and error functions \citep{MD2007} have been fit to the emission falloff. In general, methods that rely on curve-fitting or the integral of the profile are more reliable than methods that depend solely on a temperature threshold or peak location because they will be less sensitive to instrumental noise and perturbations in the H {\footnotesize I} gas density and velocity.

Assuming that the disk has a uniform number density $n$ and a spin temperature $T_s$, \citet{CRB1979} derived that the optical depth line profile close to the tangent velocity has the following form:
\begin{equation}\label{eqn:fit}
\tau(v)=f(l,n,T_s,A)\int_0^\infty\exp\left[-\frac{1}{2}\left(w^2-t\right)^2\right]dw
\end{equation}
where $t=(v-v_t)/\sigma$, $\sigma$ is the velocity dispersion, $f$ is a function that parameterizes the amplitude of the line, and $A$ is the Oort constant evaluated at the tangent point. The shape of the profile   depends only on the velocity dispersion and the tangent velocity of the gas; other parameters such as the density and shape of the rotation curve affect the amplitude. 
We convert from $\tau$ to brightness temperature $T_b$ by assuming the spin temperature is 155 K everywhere, larger than the highest $T_b$ in our data set.
\citet{RK1987} further refined eqn.~(\ref{eqn:fit}) by allowing the kinematics of the gas to deviate from circular symmetry and adding in radial motions; for the purposes of this paper, such a high level of detail is not necessary. 

The irregular shape of many profiles sometimes makes it difficult to choose which peak should be fit with eqn.~(\ref{eqn:fit}); \citet{RK1987} were forced to fit spectra interactively, determining the limits for each fit by eye. We use a more objective technique: first, we smooth each spectra with a 4 km s$^{-1}$ boxcar window. Next, we find the location of the first local maximum in the smoothed spectra with a brightness temp larger than 10 K; we call this velocity $v_{\rm max}$. We then fit the original spectrum with eqn.~(\ref{eqn:fit}) using a  Levenberg-Marquardt non-linear least squares fitting routine with three free parameters: $v_t$, amplitude, and $\sigma$. We include in this fit data from velocities in the range $v_{\rm max}-8~{\rm km~s}^{-1}\le v\le v_{\rm max}+65~{\rm km~s}^{-1}$ for QI and $v_{\rm max}-65~{\rm km~s}^{-1}\le v\le v_{\rm max}+8~{\rm km~s}^{-1}$ for QIV. These limits were optimized by minimizing the errors in parameters determined from our models, described below in \S \ref{sec:syserr}. 

\begin{figure*}
\includegraphics[angle=90,scale=.7]{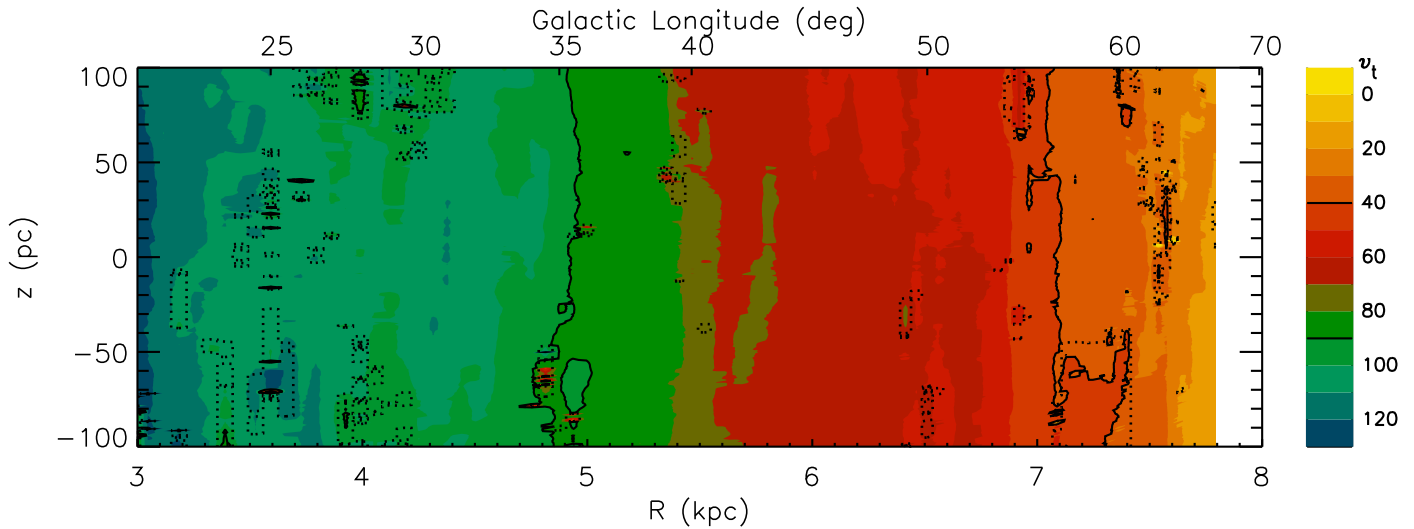}
\caption{\label{fig:vgpsvtan} Tangent velocity $v_t$ as a function of $R$ and $z$ in QI. Color contours are drawn every 10 km s$^{-1}$ and are marked on the color bar on the right in km s$^{-1}$.
The data beyond $R=7.8$ kpc are missing due to the $l$ limitations of the VGPS. Dotted lines encircle regions that have been flagged, either by the $v_t$ fitting routine or by the smoothness filter; these regions should be ignored. The $-40$ km s$^{-1}$ and $-90$ km s$^{-1}$ contours are marked with solid black lines. 
}
\end{figure*}

\begin{figure*}
\includegraphics[angle=90,scale=.7]{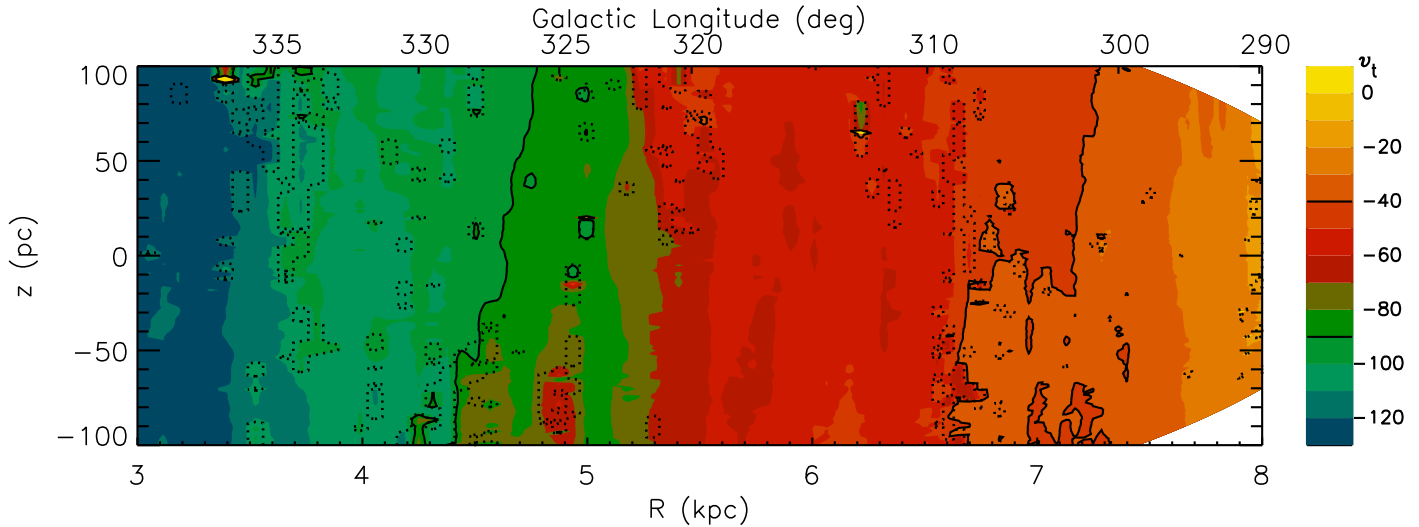}
\caption{\label{fig:sgpsvtan} Same as Fig.~\ref{fig:vgpsvtan}, but for QIV. The missing data with $R\gtrsim7.5$ kpc and $|b|>1\fdg4$ is due to the restriction on $b$ for the SGPS. The $40$ km s$^{-1}$ and $90$ km s$^{-1}$ contours are marked with solid black lines.}
\end{figure*}

The fits are generally a good match to the emission spectra. 
However, there are some exceptions. Some spectra are too irregular for the fitting routine to converge on a physically reasonable solution; an intermediate velocity cloud can fall in the range of fitted velocities, or there can be a particularly large bulk flow that distorts the shape of the spectrum. In cases like these,  $v_t$ found by the fit often falls outside the velocity boundaries of the fit. When this happens, we flag the spectra at that $(l,b)$ and remove its $v_t$ from the rest of the analysis. 0.6\% of the spectra in QI and 0.7\% of the spectra in QIV are flagged in this way.

\begin{figure*}
\includegraphics[angle=90,scale=.7]{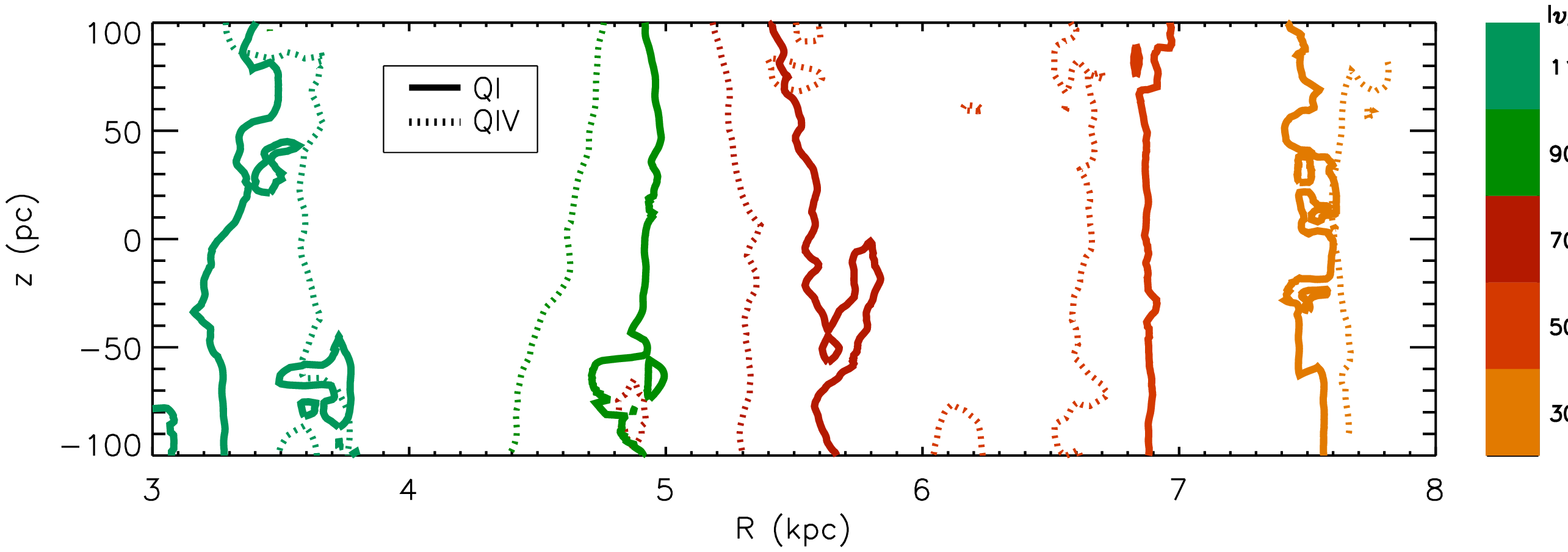}
\caption{\label{fig:combine} Contour lines for $|v_t|=30,50,70,90,110$ km s$^{-1}$ for QI (solid) and QIV (dotted). The contour values are given by the color bar.}
\end{figure*}

The map of fitted $v_t(R,z)$ is shown in Fig.~\ref{fig:vgpsvtan} for QI and Fig.~\ref{fig:sgpsvtan} for QIV. Maps on both sides of the Sun-Galactic center line show the expected large scale trend of lower $|v_t|$ at larger $R$. The $v_t$ surfaces are fairly smooth with a few small pockmarks. In a system with no change in the rotation curve with $z$, the contours in these maps would be nearly parallel to the $z$ axis, since $\cos(b)\sim1$. This is clearly not the case in many regions; for one example, examine the $-90$ km s$^{-1}$ contour in the southern map, which shows evidence for a $dv_t/dz$ term. Many contours in the $|v_t|=50$ km s$^{-1}$ to 90 km s$^{-1}$ range appear  several hundred pc closer to the Galactic center in QIV compared to QI (see Fig.~\ref{fig:combine}). This could result from a genuine asymmetry in the QI/QIV rotation curves \citep{K1962}, deviations from the axial symmetry of the potential, or from bulk flows and other peculiar motions. 

\begin{figure*}
\includegraphics[angle=0,scale=0.85]{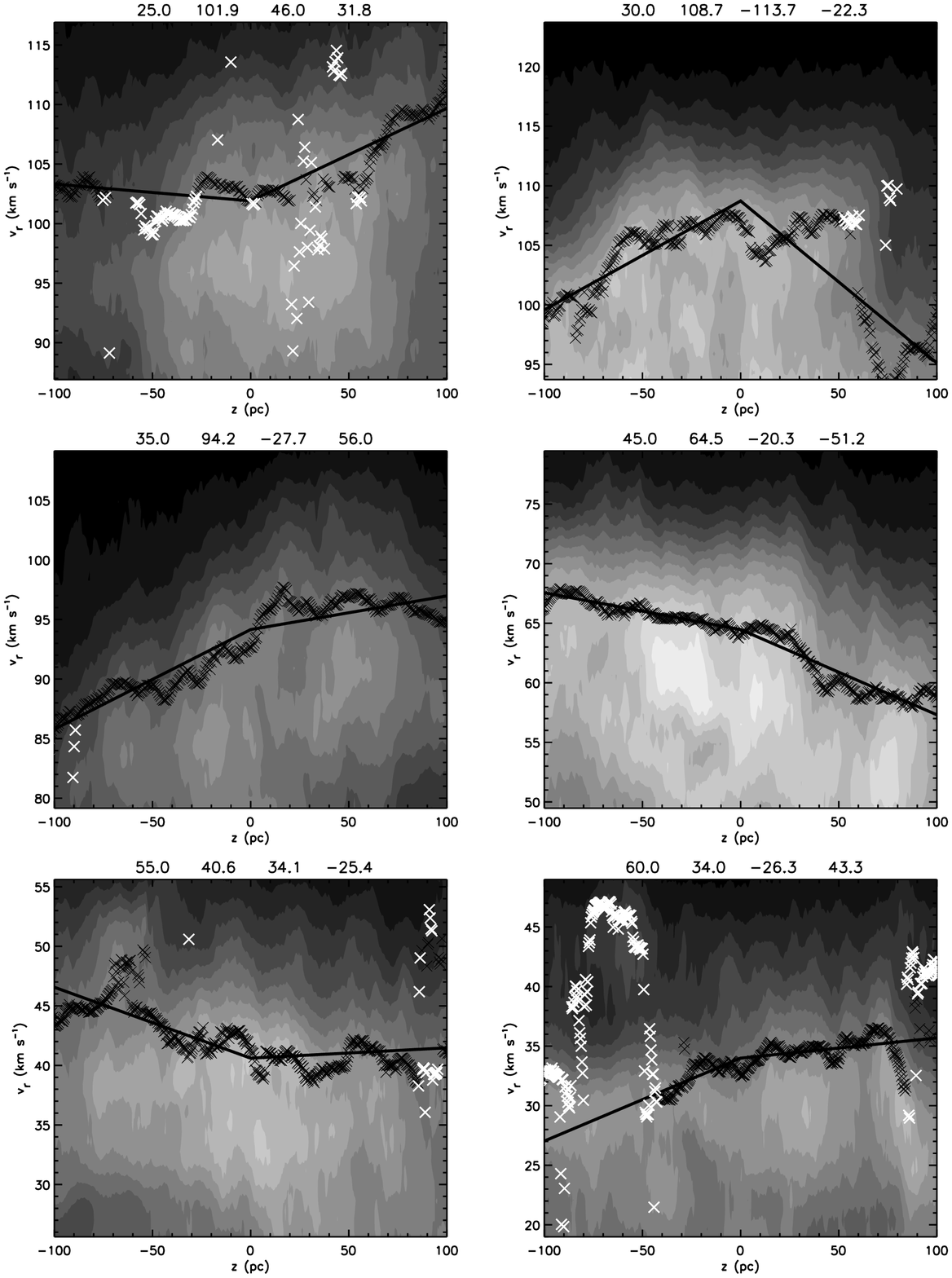}
\caption{\label{fig:vgpsplots} The measured brightness temperature as a function of $z$ and $v_r$ for several QI $l$. Contours are drawn every 10 K and lighter contours denote higher $T_b$. The x's mark $v_t$ for each $b$, and the fit to the tangent velocities is shown with a solid line. Flagged tangent velocities are marked with white x's. The fitted parameters for these $l$ from \S \ref{sec:local} are written on top of each plot in the following order:  $l$ (degrees), $v_t$ (km s$^{-1}$), $d\Theta/d|z|$ (km s$^{-1}$ kpc$^{-1}$), and $d\Theta/dz$ (km s$^{-1}$ kpc$^{-1}$).
}
\end{figure*}

\begin{figure*}
\includegraphics[angle=0,scale=0.85]{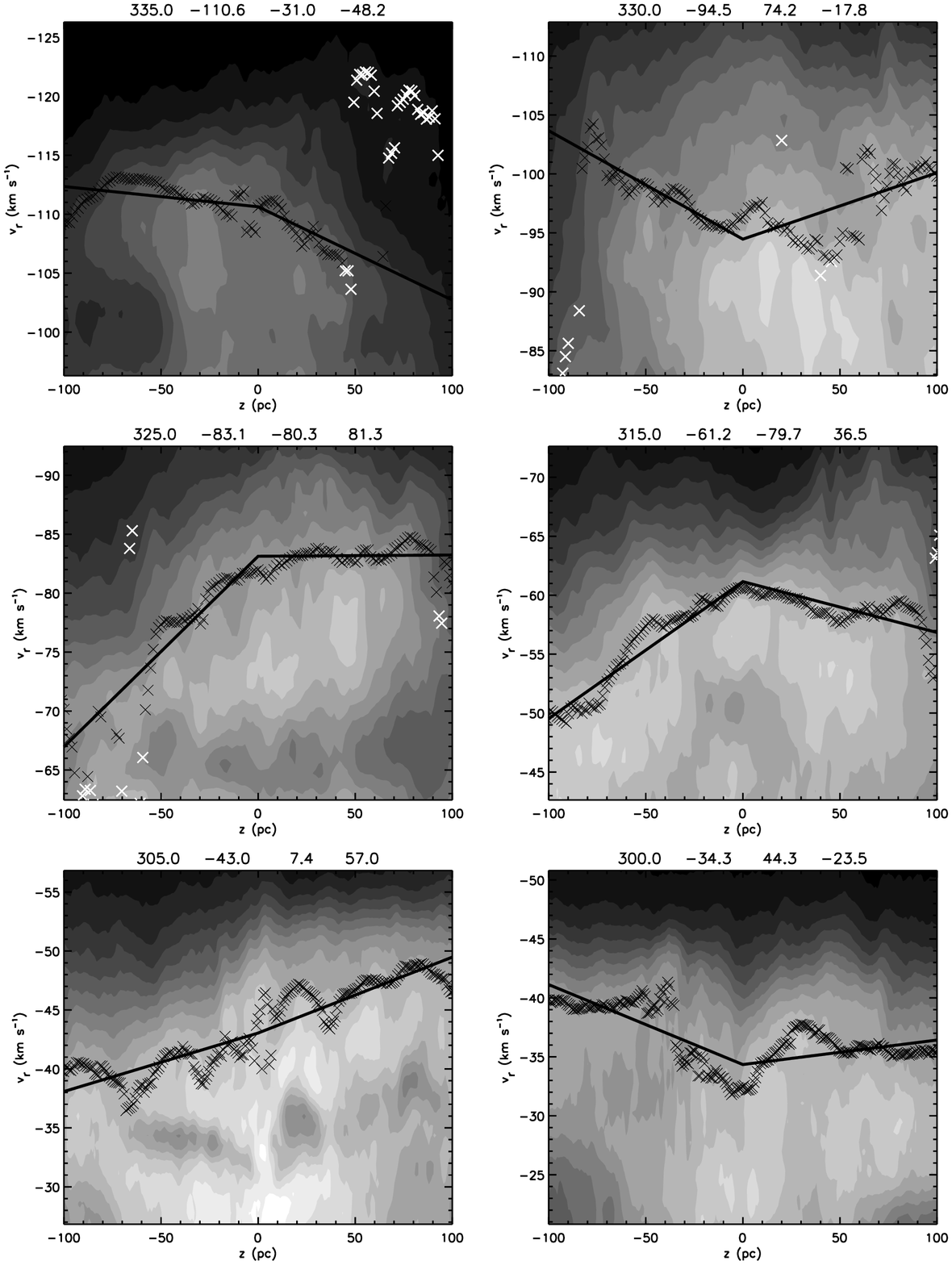}
\caption{\label{fig:sgpsplots} Same as Fig.~\ref{fig:vgpsplots}, but for QIV. Note that the velocity axis has been flipped to ease comparison with Fig.~\ref{fig:vgpsplots}.
}
\end{figure*}

In Figs.~\ref{fig:vgpsplots} and \ref{fig:sgpsplots}, we plot the tangent velocity on top of $T_b$ contours for several $l$. Note that the tangent velocity points do not simply follow a single brightness temperature contour; in general, they are at lower $T_b$ for larger $z$, because the overall amplitude of the spectra decreases away from the plane as the density falls off.  On the velocity scale of these figures the $v_t$ are quite regular, albeit with a few outliers.

\subsection{Removing $v_t$ Surface Discontinuities}
The tangent velocity surface should be a smooth function of $l$ and $b$, but the surfaces derived in \S\ref{sec:fit} have  discontinuous bumps. For example, local maxima in the spectra can result in large discontinuities in $v_t(z)$, as in the $l=335^\circ$ panel of Fig.~\ref{fig:sgpsplots} near $z\sim70$ pc. We flagged discontinuities like this one using a median-type filter and did not include them in our analysis. The filter worked as follows.

1. Calculate the function
\begin{equation}
M(l_i,b_j)=\mathrm{Median}\left[v_t(l_{i+k},b_{j+m})\right]
\end{equation}
where $-1\le k\le1$ and $-60\le m\le60$ for QI and $-30\le m\le30$ for QIV. In other words, $M$ is the median of all the points inside a rectangular box in $(l,b)$ space with a width of $3\times121$ elements for QI and $3\times61$ elements for QIV; the allowed ranges of $k$ and $m$ set the size of the median filter in the $l$ and $b$ dimensions, respectively. The box encompassed  more elements in $b$ than in $l$ because we already reduced the number of elements in $l$ as discussed in \S \ref{sec:data}. Twice as many points were included in $b$ in QI because of the higher resolution of the VGPS compared to the SGPS. The stepsize in $l$ for both quadrants was $0\fdg5$, while in $b$ it was $0\fdg005$ for QI and $0\fdg011$ for QIV. Points that were flagged due to poor fits in \S \ref{sec:fit} were not included in the median calculation.

2. If $|v_t(l_i,b_j)-M(l_i,b_j)|>5$ km s$^{-1}$ then flag $v_t(l_i,b_j)$ and do not include it in our analysis. This threshold eliminated most of the bumps that were inconsistent in adjacent longitudes. 

Flagged data points for a few $l$ are shown with white x's in QI (Fig.~\ref{fig:vgpsplots}) and QIV (Fig.~\ref{fig:sgpsplots}). 4.2\% of the tangent velocities in QI were flagged by this filter, as were 6.9\% of the velocities in QIV. This filter generally flagged the $v_t$ points where the tangent velocity varies erratically. 

Discontinuities that persist over a large angular scale were not removed by this filter. In particular, a large discontinuity is located between $l=57^\circ$ and $60\fdg5$ with $-50~\mathrm{pc}\lesssim z\lesssim -100$ pc (see Fig.~\ref{fig:vgpsplots}, $l=60^\circ$). We flagged this region by hand and removed it from the analysis.

\section{Global Rotation Curve Fit}
\label{sec:global}

To model the global rotation curve, we extended a linear rotation curve explored in \citet{FBS1989} by adding two vertical derivative terms:
\begin{equation}
\Theta(R,z)=a_1\Theta_0+a_2\frac{\Theta_0}{R_0}R+\frac{d\Theta}{d|z|}|z|+\frac{d\Theta}{dz}z.
\end{equation}
The global rolling motion is given by $d\Theta/dz$, and the global falloff from the plane by $d\Theta/d|z|$. Throughout this paper, we call these the rolling parameter and the falloff parameter, respectively.
 Using eqn.~(\ref{eqn:tan}), we then have an equation for the observed tangent velocities:
\begin{eqnarray}\label{eqn:global}
v_t(l,b)=\cos(b)\Bigl[&\pm& a_1\Theta_0 \pm a_2\frac{\Theta_0}{R_0}R \nonumber\\ &\pm& \frac{d\Theta}{d|z|}|z|\pm \frac{d\Theta}{dz}z-\Theta_0\sin(l)\Bigr],
\end{eqnarray}
where the ``$+$'' is for QI and the ``$-$'' for QIV.
Alternatively, we could  have replaced the third term with $z^2 d^2\Theta/dz^2$, but this choice did not affect our results. 

Since points are unequally spaced in $(R,z)$ space, performing a simple linear least squares fit would unevenly weight some $R$ more than others. Also, the stepsize in $b$ in QI is smaller than in QIV, so QI would have a larger influence on a combined fit. Furthermore, although there are more than a hundred spectra at each longitude, many of the derived tangent velocities are not independent. Bulk flows and other peculiar motions are responsible for most of the deviations from circular motion; these motions have a typical velocity scale of 8 km s$^{-1}$. Small regions can vary from the local rotational speed by 20-30 km s$^{-1}$ \citep{BB1993}. The physical scale of these bulk flows is difficult to determine, but we estimate there are approximately 10 independent data points with $|z|\le100$ pc for each $l$.

We construct an interpolated data set with a $z$ grid given by $-90,-70,-50,\ldots,90$ pc but leaving unchanged the spacing in $l$. We flagged points in the interpolated grid if either of the two points adjacent in $b$ in the original grid were flagged.
We performed a linear least squares fit to the interpolated data within the range 3 kpc $\le R \le$ 8 kpc and $|z|\le100$ pc. The results for the QI, QIV, and combined data set  are shown in Table \ref{tab:global}, along with the formal errors in the fit. We emphasize that these rotation curve parameters are valid only inside $|z|\le100$ pc; we have not constrained the falloff outside of this region.

What is the effect of the bulk flows on the values of the fitted parameters? In short, quantifying the impact of bulk flows requires a self-consistent simulation of gas dynamics in the Milky Way disk, which is beyond the scope of this investigation. We have assumed that these random motions do not have a strong effect on the values of the fitted parameters, other than to correlate nearby data points.

Are there places in the inner Galaxy where this analysis is inappropriate?
Noncircular orbits can result from the presence of a stellar bar; this would invalidate an assumption that led to eqn.~(\ref{eqn:tan}).
A recent survey of mid-infrared sources toward the inner Galaxy detected a QI-QIV asymmetry in star counts for $l\lesssim30^\circ$;
this corresponds to a bar radius of $\sim$4.4 kpc \citep{BCBIM2005}. We also performed a global fit restricted to spectra with 4.4 kpc $\le R\le$ 8 kpc, but this did not strongly affect the derived parameters. We defer an in depth discussion of the physical interpretation of the parameter values until \S \ref{sec:local}.

\begin{deluxetable*}{cccccc}
\tablecaption{Global Least-Squares Fit Parameters\label{tab:global}}
\tablehead{
\colhead{Data}&\colhead{$a_1$}&\colhead{$a_2$} &\colhead{$d\Theta/d|z|$}&\colhead{$d\Theta/d|z|_{\rm corr}$}&\colhead{$d\Theta/dz$}}
\startdata
QI&0.855$\pm$0.004&0.209$\pm$0.006&-19$\pm$7&-26$\pm$8&-8$\pm$4\\
QIV&0.829$\pm$0.004&0.221$\pm$0.005&-12$\pm$7&-16$\pm$8&9$\pm$3\\
Combined&0.844$\pm$0.003&0.212$\pm$0.004&-16$\pm$5&-22$\pm$6&1$\pm$2
\enddata
\tablecomments{The global rotation curve assumes $R_0=8.5$ kpc and $\Theta_0=220$ km s$^{-1}$. The parameters $a_1$ and $a_2$ are unitless, $d\Theta/d|z|$ and $d\Theta/dz$ are in km s$^{-1}$ kpc$^{-1}$, and $d\Theta/d|z|_{\rm corr}$ is the falloff parameter corrected for the systematic error discussed in \S \ref{sec:globmodel}.}
\end{deluxetable*}

\section{Local Rotation Curve\\ Parameter Determinations}
\label{sec:local}
Do the vertical derivatives of the rotation curve vary as a function of $R$? We investigated  by fitting the vertical derivatives at each $l$ independently.
We approximated the rotation curve by the first three terms of its Taylor expansion, 
\begin{equation}\label{eqn:rotmodel}
\Theta(R,z)\approx\Theta(R,0)+|z|\frac{d\Theta}{d\left|z\right|}\Big|_{z=0}+z\frac{d\Theta}{dz}\Big|_{z=0}.
\end{equation}
We then wrote the equation for the tangent velocity in terms of the $b=0$ solution to eqn.~(\ref{eqn:tan}), $v_{t,0}$:
\begin{equation} 
v_t(b)=\cos(b)\left[v_{t,0}\pm|z|\frac{d\Theta}{d\left|z\right|}\Big|_{z=0}\pm z\frac{d\Theta}{dz}\Big|_{z=0}\right],\label{eqn:tangentv}
\end{equation}
where again the ``$+$'' is for QI and the ``$-$'' for QIV.
Note that this equation has the observables $v_t(b)$ and three unknowns: $v_{t,0}$, $d\Theta/d|z|$, and $d\Theta/dz$.
For each $l$ in our sample, we determined the three rotation curve parameters in eqn.~(\ref{eqn:tangentv}) using a linear least squares fit. Several fit samples in QI (Fig.~\ref{fig:vgpsplots}) and QIV (Fig.~\ref{fig:sgpsplots}) are shown as solid lines on top of the  $T_b$ contours and $v_t$ fits. 

Again we have the problem of bulk flows correlating the tangent velocities at nearby points. Ideally, we would proceed by calculating the correlation matrix to use in the fit, but the physical scale of the flows is not well determined. Instead, we adopt an ad hoc solution: calculate the formal error in the fit assuming the data points are independent, and then scale up the errors by assuming there are only ten independent data points between $-100$ pc $\le z\le 100$ pc, i.e.~we multiplied the formal errors by $\sqrt{\mathrm{number~of~points}/10}$.

We now discuss each of the three fitted parameters.

\subsection{Rotation Curve at $z=0$}

\begin{figure}
\includegraphics[angle=90,scale=.35]{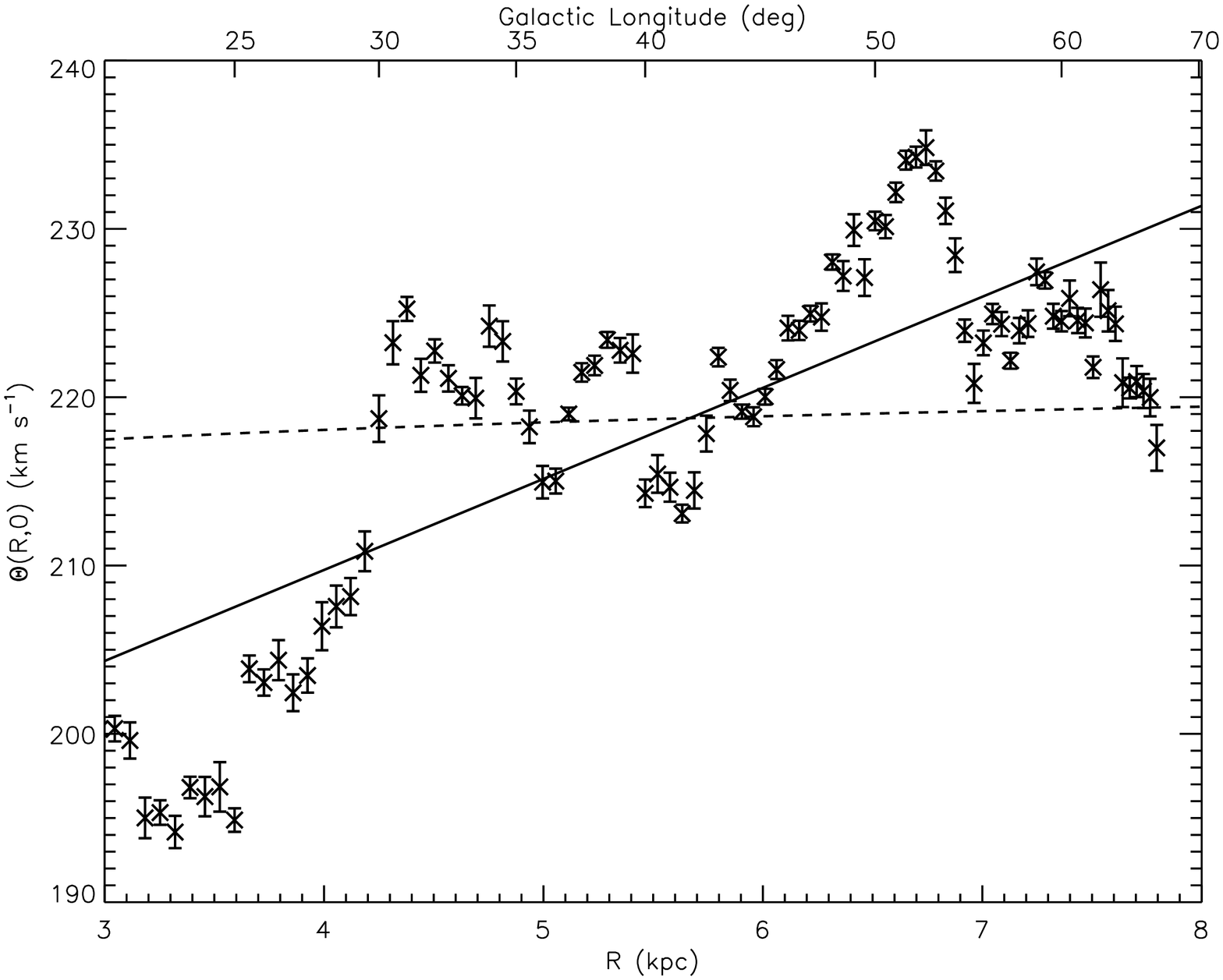}
\caption{\label{fig:vgpsrot} The rotation curve $\Theta(R,0)$ as measured at the tangent points in QI. The solid line is the rotation curve fit to the QI data from \S \ref{sec:global}, and the dashed line is the fit to QI data from \citet{BB1993}. }
\end{figure}

\begin{figure}
\includegraphics[angle=90,scale=.35]{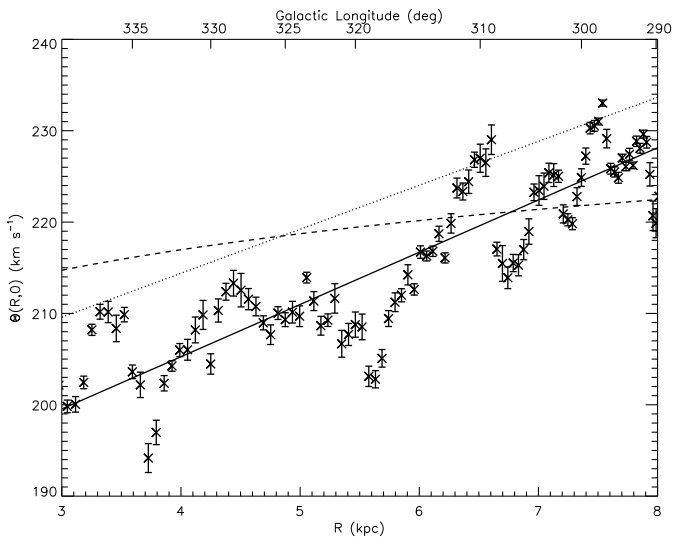}
\caption{\label{fig:sgpsrot} Same as Fig.~\ref{fig:vgpsrot} but for QIV. The solid line is the rotation curve fit to the QIV data from \S \ref{sec:global}, the dashed line is the fit to QIV data from \citet{BB1993}, and the dotted line is a fit to the SGPS data from \citet{MD2007}. }
\end{figure}

In Figs.~\ref{fig:vgpsrot} and \ref{fig:sgpsrot} we plot the parameter $\Theta(R,0)$ in QI and QIV, respectively. Note that, for the purposes of this plot, we have assumed $\Theta_0=220$ km s$^{-1}$ with no error, but  the determination of $v_{t,0}$ does not depend on the value of $\Theta_0$. The SGPS for $R\gtrsim7.5$ kpc does not cover the full range of $|z|<100$ pc because of the limited survey extent in $b$. For these longitudes, we remove the fits  of the two derivatives from our analysis, but keep the easier to determine $\Theta(R,0)$.
In both QI and QIV the rotation curve rises as $R$ increases, consistent with previous measurements of the inner Galaxy rotation curve \citep{KMW1954}.
On top of the $\Theta(R,0)$ fit points, we plot the global fit to the rotation curve from \S\ref{sec:global}. The fit traces the measured rotation curve well, reinforcing the conclusions of our global analysis.

\begin{figure}
\includegraphics[angle=90,scale=.35]{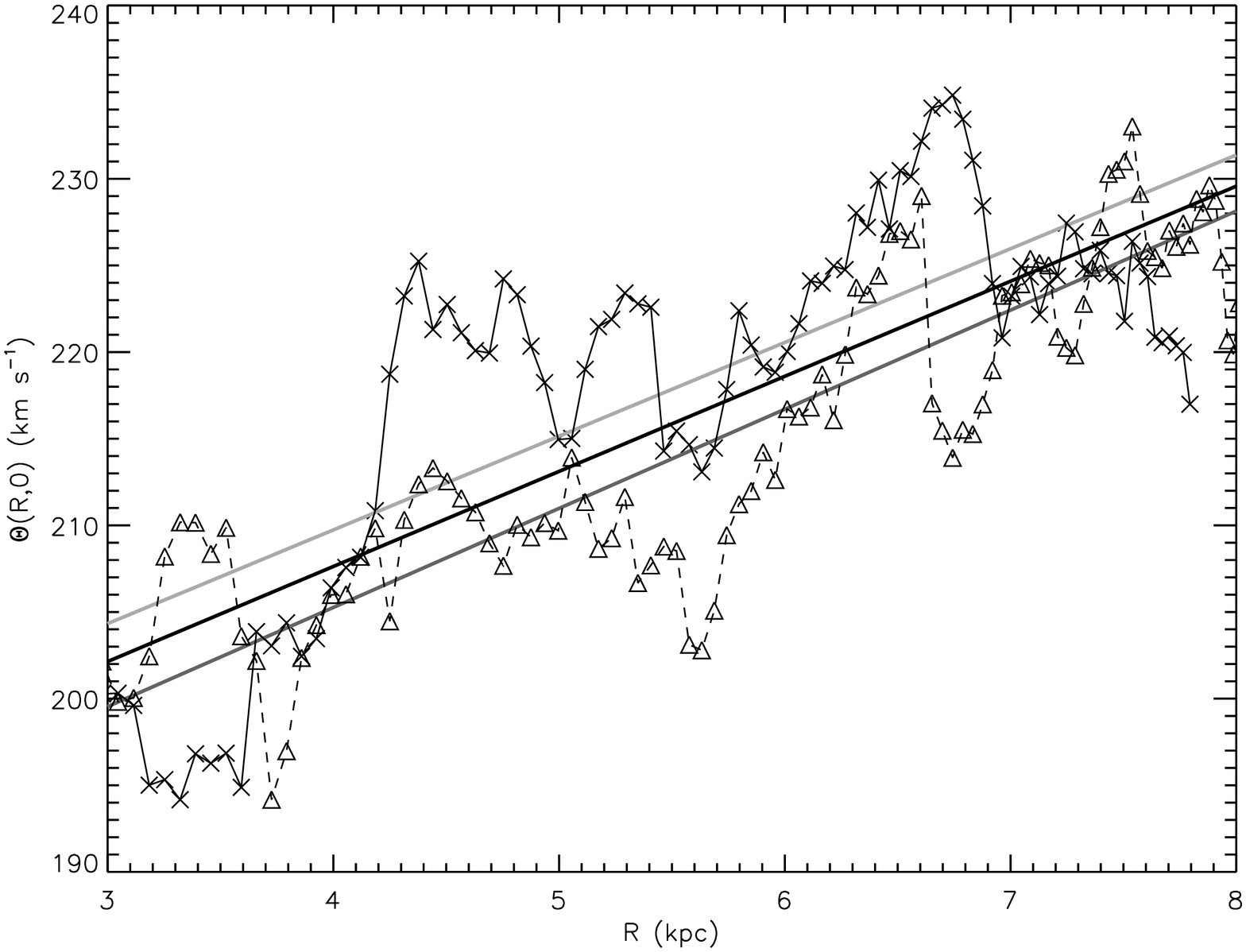}
\caption{\label{fig:combrot} The rotation curve $\Theta(R,0)$ as measured at the tangent points in the inner Galaxy. The x's connected by a solid line are the QI values; the triangles connected with a dashed line are the QIV ones. The rotation curve from the fit to the QI and QIV data (Table \ref{tab:global}) is shown with the heavy solid line, the QI fit is shown by the light grey line, and the QIV fit is shown by the dark grey line.
}
\end{figure} 

Our fit to the QIV  rotation curve is approximately 7 km s$^{-1}$ lower than the one determined in \citet{MD2007}. We suspect this discrepancy is due to the differences in $v_t$ fitting methods. In both quadrants, our data are not well matched by the fit  from \citet{BB1993}. However, data from the outer Galaxy was also used to constrain their fit, so this is not surprising. 

The QI and QIV rotation curves are shown on the same plot in Fig.~\ref{fig:combrot}. 
The asymmetry in the northern and southern rotation curves is quite similar in shape to that measured by \citet{K1964}, who observed that the QI curve is larger than the QIV curve between $4.2$ kpc $\le R\le 6.8$ kpc.
 Compared to this previous work,  the magnitude of our curves is different due to the change in the preferred values for Galactic constants. More recent work shows the same asymmetry \citep{BS1991}.

Bumps in the inner Galaxy rotation curve have long been interpreted as the locations where spiral arms cross the tangent circle \citep{Y1969,B1971}. We will not speculate on the meaning of the bumps, other than to say they are about 10 km s$^{-1}$ in amplitude, as expected from density wave theory \citep[see][]{B1971}.

\subsection{Falloff Parameter}

\begin{figure}
\includegraphics[angle=90,scale=.35]{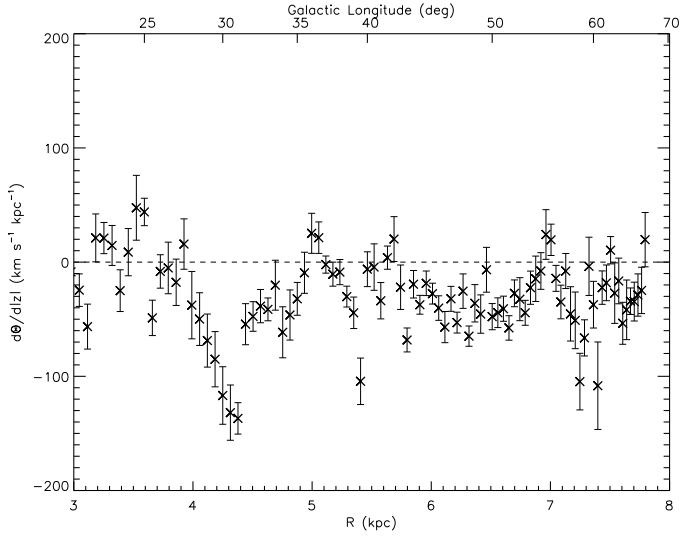}
\caption{\label{fig:vgpsdtdz} The radial dependence of $d\Theta/d|z|$ as measured at the tangent points in QI. Error bars  are from the adjusted error in the fit (\S \ref{sec:local}). }
\end{figure}

\begin{figure}
\includegraphics[angle=90,scale=.35]{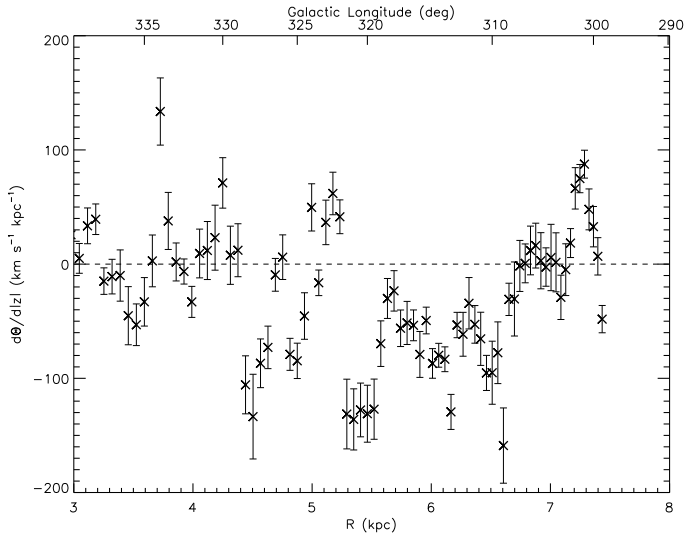}
\caption{\label{fig:sgpsdtdz} Same as Fig.~\ref{fig:vgpsdtdz} but for QIV. }
\end{figure}

The falloff parameter $d\Theta/d|z|$ is shown for QI in Fig.~\ref{fig:vgpsdtdz} and QIV in Fig.~\ref{fig:sgpsdtdz}. In QI, the falloff appears  centered near $-30$ km s$^{-1}$ kpc$^{-1}$. We include a correction for a $\sim5$ km s$^{-1}$ kpc$^{-1}$ systematic error due to the falloff of gas density with $z$ discussed in \S\ref{sec:model}. The falloff parameter is roughly constant with radius, although small oscillations are apparent throughout, and a large jump is seen near $R\sim4.5$ kpc. In comparison, the falloff in QIV is much less regular. The points are also centered around $-30$ km s$^{-1}$ kpc$^{-1}$, but the scatter is much larger and their values are less correlated from one point to the next. There also is a large group of points near $R\sim7$ kpc where $d\Theta/d|z|>0$ km s$^{-1}$ kpc$^{-1}$; it is unclear what the significance of this group of points is, and also why the QI and QIV plots have such a different appearance in detail. 

Overall, the local analysis again confirms the global fit because they both point towards a falloff parameter with magnitude approximately $-20$ km s$^{-1}$ kpc$^{-1}$. The roughly constant value of $d\Theta/d|z|$ in QI consistent with the global fit from \S\ref{sec:global} leads us to conclude that the falloff is a global feature of the inner Galaxy rotation curve (at least within 100 pc of the plane), and is not limited to any one region.

\subsection{Rolling Parameter}

\begin{figure}[t]
\includegraphics[angle=90,scale=.35]{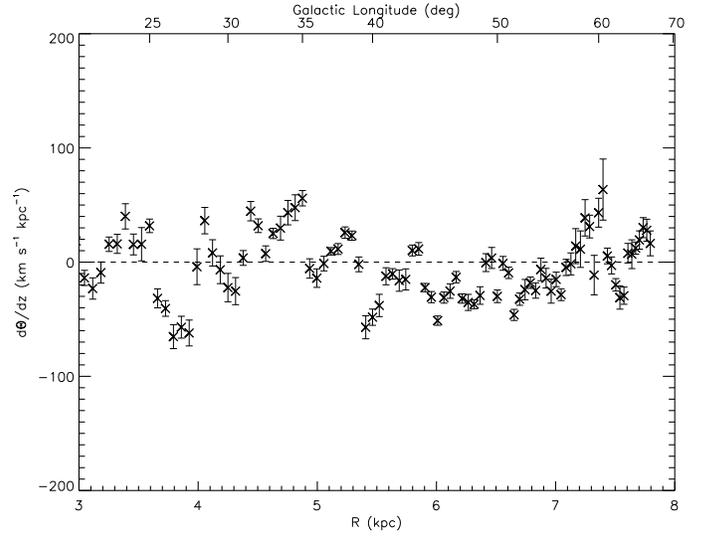}
\caption{\label{fig:vgpsslope} The radial dependence of $d\Theta/dz$ as measured at the tangent points in QI. Error bars  are from the adjusted error in the fit (\S \ref{sec:local}). }
\end{figure}

\begin{figure}[!t]
\includegraphics[angle=90,scale=.35]{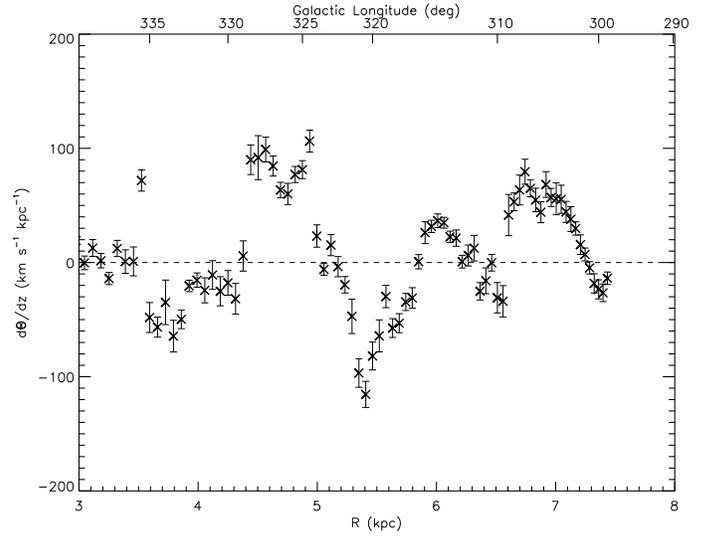}
\caption{\label{fig:sgpsslope} Same as Fig.~\ref{fig:vgpsslope} but for QIV. }
\end{figure}

The rolling parameter $d\Theta/dz$ is shown for QI in Fig.~\ref{fig:vgpsslope} and QIV in Fig.~\ref{fig:sgpsslope}. In both these figures, nearby points are highly correlated and the curves have several peaks and valleys. Both plots appear to be centered near zero velocity. 

There is an asymmetry between the QI and QIV values of the rolling parameter in that the QIV data have higher amplitude oscillations than those in QI. The high level of rolling motion in the QIV tangent circle may be the reason the QIV falloff parameter has high scatter compared to that of QI. Though the two derivatives are nearly orthogonal, the falloff may be easier to detect in a region with quiescent H {\footnotesize I} and the QIV tangent circle clearly has a good deal of kinetic energy due to rolling motions. The correlation length of the data in QIV also appears to be larger than that of QI.

It is difficult to compare the rolling motions in these figures to those observed previously by \citet{FS1985}. The authors claim typical values of $\pm20$ km s$^{-1}$ kpc$^{-1}$ and values reaching $-75$ km s$^{-1}$ kpc$^{-1}$ in localized regions. They determined these numbers by tracing the tilt in $T_b$ contours specifically in spiral arms, which are not directly comparable to our calculations. Nevertheless,  the two methods of measuring rolling motions satisfyingly produce slopes of the same order-of-magnitude. 

Because of the oscillatory nature of the rolling motions in both quadrants, as well as the global fit consistent with zero slope, we conclude that the rolling motions average to zero over the inner Galaxy (again, within 100 pc of the plane). However, local fits demonstrate that slopes as large as $\pm$100 km s$^{-1}$ kpc$^{-1}$  appear in localized regions. 

\section{Fit Variations and Systematic Errors}\label{sec:fitvar}
\subsection{Fit Variations}

How dependent are the parameter values on the method of fitting $v_t$? We refit all the data using the integral method of \citet{SB1966}, and find that the $v_t$ are generally $\sim$5-10 km s$^{-1}$ larger in magnitude than those determined with the curve fitting method, as predicted by \citet{BG1978}. This results in a rotation curve with higher velocities. If we run this data set through our global fitting routine, the falloff parameter for the QI and QIV combined data is $-12\pm4$ km s$^{-1}$ kpc$^{-1}$, within $1\sigma$ of the result from \S \ref{sec:global} using the  \citet{CRB1979} $v_t$ fitting method. The rolling parameter has a global value of $1\pm2$ km s$^{-1}$ kpc$^{-1}$, essentially a match.
 We conclude that our use of the analytic model fitting method does not  bias the appearance of the vertical derivatives.

We have also tried replacing $z$ in eqn.~(\ref{eqn:tangentv}) with $(z-z_m)$, where $z_m$ is the center of a Gaussian fit to the brightness temperature as a function of $z$ for a given $R$ (in other words, the midplane of the gas distribution). The midplane  can be displaced by up to 50 pc from $z=0$ in the inner Galaxy, though smaller values are more typical \citep{M1995}. We ran the local fit to the QI data with this modification, and saw little change in the values of the two derivatives.

The Sun is located about 15 pc above the Galactic plane (\citealt{FBD1994} and references therein, \citealt{C1995}, \citealt{NBCB1997}, \citealt{BGS1997}, \citealt{MBM1985}). We checked if this influenced our results by recalculating the vertical derivatives with an adjusted $z$ coordinate, $z_{\rm adj}=z+15~{\rm pc}-R_0\cos(l)\tan(0.1^\circ)$. The change in $a_1$, $a_2$, and $d\Theta/d|z|$ were all within the 1$\sigma$ errors, but the combined value for $d\Theta/dz$ became $6\pm2$ km$^{-1}$ kpc$^{-1}$.

\subsection{Systematic Errors}\label{sec:syserr}
\subsubsection{Testing Global Fit Accuracy with Models}\label{sec:globmodel}

For the purpose of testing our global rotation curve parameter determinations (\S\ref{sec:global}), we created simple models of Galactic H {\footnotesize I} emission  with the same $l$ and $b$ spacing and $v_r$ resolution as the SGPS. We assumed a constant H {\footnotesize I} surface density of 4 M$_\odot$ pc$^{-2}$ \citep{BG1978}  and a two-component  gas,  80\% with a velocity dispersion of 8 km s$^{-1}$ and the remainder with a dispersion of 35 km s$^{-1}$ \citep{KF1985}. 
The  gas in the model  is assumed to follow a sech$^2$ distribution in $z$ \citep{S1942} with a scale height of 120 pc \citep{L1984}, and for simplicity we did not change the proportions of the different velocity dispersion components as a function of height. We included optical depth effects using a spin temperature of 155 K. We constructed several different models for the change in the rotation curve as a function of $R$ and $z$ (Table \ref{tab:globmodel}). We did not add random noise to the models because we are interested in estimating the systematic uncertainty solely due to the falloff in gas density with $|z|$.  

After reducing the size of the model data using the median method described in \S \ref{sec:data} and fitting the $v_t$ as in \S \ref{sec:fit}, we performed the linear least squares fit described in \S \ref{sec:global} to recover the parameters of $\Theta(R,z)$ in each case. Table \ref{tab:globmodel} shows that 
 the input parameters were essentially recovered, suggesting a small systematic error in all parameters except  $d\Theta/d|z|$, where the error  is generally  $4\pm1$ km s$^{-1}$ kpc$^{-1}$. This error occurs because we fit $v_t$ at latitudes where $b\ne 0$, and the assumption in eqn.~(\ref{eqn:fit}) that $n$ is constant along the line of sight is not justified. On average, points further along the line of sight will have smaller $n$ than more nearby points, since the far points are at a higher $z$ and the gas is generally densest at $b=0$.
 This varying gas density causes small inaccuracies in the fitted $v_t$ which translate to a systematic error in the falloff parameter. 
  The surface density of H {\footnotesize I}  plays a role in setting the magnitude of this effect. We correct for this systematic error in the column labeled $d\Theta/d|z|_{\rm corr}$ in Table \ref{tab:global}.  The systematic errors in $a_1$ and $a_2$ are sometimes significant, but we do not correct for them here. We also modeled QI with $a_1=0.82$, $a_2=0.22$, $d\Theta/d|z|=-30$ km s$^{-1}$ kpc$^{-1}$, and $d\Theta/dz=0$, similar to the measurements in \S \ref{sec:global}, and we found the error in the falloff parameter was $7$ km s$^{-1}$ kpc$^{-1}$. When the QI and QIV data sets were combined, the error in the falloff parameter was $6$ km s$^{-1}$ kpc$^{-1}$. From our experience with the QIV models, we assigned an error in these systematic corrections of $\pm1$ km s$^{-1}$ kpc$^{-1}$.

\begin{deluxetable*}{ccccccc|cccc}
\tablecaption{Global Model and Fit Parameters\label{tab:globmodel}}
\tablehead{\multicolumn{7}{c|}{Model}&\multicolumn{4}{c}{Fit}\\
\colhead{$a_1$}&\colhead{$a_2$}&\colhead{$d\Theta/d|z|$}&\colhead{$d\Theta/dz$} &\colhead{$\sigma_1$}&\colhead{$\sigma_2$}&\multicolumn{1}{c|}{$\sigma_3$}&\colhead{$\Delta a_1$}&\colhead{$\Delta a_2$}&\colhead{$\Delta d\Theta/d|z|$}&\colhead{$\Delta d\Theta/dz$}}
\startdata
1.00&0.00&-30&-30&8&35&\nodata&-0.009&0.007&4&1\\ %sgpsfull
1.00&0.00&30&30&8&35&\nodata&-0.009&0.007&4&1\\ %sgpsfull2
0.82&0.22&-30&30&8&35&\nodata&-0.014&0.013&5&1\\ %sgpsfull3
0.82&0.22&-30&-30&8&35&\nodata&-0.014&0.013&5&1\\ %sgpsfull4
0.82&0.22&30&0&8&35&\nodata&-0.013&0.012&4&0\\ %sgpsfull5
0.82&0.22&0&0&8&35&\nodata&-0.013&0.012&4&0\\ %sgpsfull6
0.82&0.22&-30&0&8&35&\nodata&-0.014&0.013&5&0 %sgpsfull8
\enddata
\tablecomments{The derivatives $d\Theta/d|z|$ and $d\Theta/dz$ are given in km s$^{-1}$ kpc$^{-1}$. The parameters $a_1$ and $a_2$ from eqn.~(\ref{eqn:global}) are unitless. Velocity dispersions $\sigma$ are given in km s$^{-1}$. The systematic errors given in the fit column are defined as $\Delta x=x_{\rm fit}-x_{\rm model}$.}
\end{deluxetable*}

\subsubsection{Testing Local Fit Accuracy with Models}\label{sec:model}

To test our local parameter determinations (\S\ref{sec:local}), we created  simple models of Galactic H {\footnotesize I} emission at $l=317\fdg5$ ($R$=6.2 kpc) with the same $l$ and $b$ spacing and $v_r$ resolution as the SGPS. We used the same gas characteristics as our previous models in \S \ref{sec:globmodel}, except we tried several different velocity dispersion combinations. The basic model has a two component gas, 80\% with a velocity dispersion of 8 km s$^{-1}$ and the remainder with a dispersion of 35 km s$^{-1}$ \citep{KF1985}.
 A recent analysis has suggested a three component model with 56\% of the gas with a velocity dispersion of  6 km s$^{-1}$, 38\% with 12 km s$^{-1}$, and 6\% with 26 km s$^{-1}$ gas \citep{MD2007}; we construct a model with these parameters as well. For completeness, we also construct a model made solely of  8 km s$^{-1}$ gas \citep{B1976}. We made models with various values of velocity dispersions and vertical derivatives.

We performed the fit described in \S \ref{sec:fit} to determine $v_t(l=317\fdg5,b)$ for these models, and tried to recover the parameters of $\Theta(R=6.2~\mathrm{kpc},z)$ in each case (Table \ref{table:models}). 
 The input parameters are essentially recovered, suggesting systematic errors of $\sim$ 1 km s$^{-1}$  in $v_{t,0}$ and $\sim$ 5 km s$^{-1}$ kpc $^{-1}$ in the falloff again due to the change in gas density with $z$. 
 For models with more than one velocity dispersion, it is possible to perform a more complicated fit by summing multiple versions of eqn.~(\ref{eqn:fit}), one for each component, but  we found that this does not improve the accuracy of the fitted values of $\Theta$ and its derivatives.

Although the systematic errors in the falloff parameter local analysis do not vary as a function of the gas velocity dispersion, they do vary as a function of $R$. To correct for this effect, we used the global model with parameters $a_1=0.82$, $a_2=0.22$, $d\Theta/d|z|=-30$ km s$^{-1}$ kpc$^{-1}$, and $d\Theta/dz=0$ km s$^{-1}$ kpc$^{-1}$ and ran it through our local analysis. The routine returned errors in $d\Theta/d|z|$ ranging from $\sim-1$ km s$^{-1}$ kpc$^{-1}$ at $R=3$ kpc  to $\sim-12$ km s$^{-1}$ kpc$^{-1}$ at $R=8$ kpc. These magnitude of these errors is expected from the errors in the global analysis. We corrected the local measurements of the falloff parameter in Figs. \ref{fig:vgpsdtdz} and \ref{fig:sgpsdtdz}.

\begin{deluxetable*}{cccccc|ccc}
\tablecaption{Local Model and Fit Parameters\label{table:models}}
\tablehead{\multicolumn{6}{c|}{Model}&\multicolumn{3}{c}{Fit}\\
\colhead{$v_{t,0}$}&\colhead{$d\Theta/d|z|$}&\colhead{$d\Theta/dz$} &\colhead{$\sigma_1$}&\colhead{$\sigma_2$}&\multicolumn{1}{c|}{$\sigma_3$}&\colhead{$\Delta v_{t,0}$}&\colhead{$\Delta d\Theta/d|z|$}&\colhead{$\Delta d\Theta/dz$}}
\startdata
-71&30 & 0& 8&35&\nodata&1&5&0\\ %sgpsc1
-71&0 &0& 8&35&\nodata&1&5&0\\ %.01 sgpsc2
-71&-30 &0&8&35&\nodata&1&5&0\\ %sgpsc3
-71&-60 &0&8&35&\nodata&1&5&0\\ %sgpsc4
-71&-90 &0&8&35&\nodata&1&5&0\\ %sgpsc5
-71&-30&0&8&\nodata&\nodata&1&0&0\\ %sgpsm301
-71&-30&0&6&12&26&1&2&0\\ %sgpsm30
-71&-30&-30&8&35&\nodata&1&5&0\\ %sgpsc6
-71&-30&30&8&35&\nodata&1&5&0 %sgpsc7
\enddata
\tablecomments{Tangent velocities $v_{t,0}$ and velocity dispersions $\sigma$ are given in km s$^{-1}$. The derivatives $d\Theta/d|z|$ and $d\Theta/dz$ are given in km s$^{-1}$ kpc$^{-1}$. The systematic errors given in the fit column are defined as $\Delta x=x_{\rm fit}-x_{\rm model}$.}
\end{deluxetable*}

\section{Discussion}

Averaging spectra over $b$, as other authors have done \citep{M1995,MD2007}, can result in errors of $\sim$1-2 km s$^{-1}$ in the determination of $\Theta(R,0)$ because the rotation speed changes with height. This effect is  more harmful in surveys that are limited in $b$ (compared to a survey limited in $z$), because their range of $z$ varies as a function of $l$. A systematic error can therefore be introduced. A similar problem occurs if the data do not extend to the same distance in $b$ on either side of $z=0$. If it is necessary that spectra be averaged to improve signal to noise, we recommend averaging over a set limited in $z$ rather than  one limited in $b$.  

Is the magnitude of the falloff parameter surprising? What  change in the rotation speed is expected due the shape of the gravitational potential? We assumed the potential is dominated by the stellar disk contribution. At some locations in the disk this assumption will not hold, but spherical components like the bulge and the halo result in negligibly small $d\Theta/d|z|$ at $|z|=100$ pc. Next, we assumed that the gas travels in non-intersecting cylindrical orbits, with the rotation speed exactly balancing the radially inward gravitational force, $g_R$, resulting in the relation $\Theta(R,z)=\sqrt{-Rg_R}$. For an exponential vertical distribution of stellar mass, \citet{PPCL2002} found that
\begin{equation}
g_R(R,z)=-2\pi G \int_0^\infty \beta S_0(\beta) J_1(R\beta)\frac{\frac{\beta}{\alpha}e^{-\alpha|z|}-e^{-\beta|z|}}{\left(\frac{\beta}{\alpha}\right)^2-1} d\beta
\end{equation}
where
\begin{equation}
S_0(\beta)=\int_0^\infty r J_0(\beta r)\Sigma(r) dr.
\end{equation}
$J_0$ and $J_1$ are the zeroth and first order Bessel functions, $\Sigma(R)$ is the stellar surface density, and $\alpha$ is the inverse of the stellar scale height $h$. We solved these equations using $h=335$ pc, $\Sigma(R)=\Sigma_0\exp(-R/R_d)$, $R_d=2.5$ kpc, and $\Sigma(R_0)=54$ M$_\odot$ pc$^{-2}$. Within the range 3 kpc $\le R\le8$ kpc, this model predicts $|d\Theta/d|z|| \lesssim 5$ km s$^{-1}$ kpc$^{-1}$, too small to explain the magnitude of the observed effect by about a factor of four. Some other process besides simple gravitational physics must be at work.

In Table \ref{table:summ} we summarize measurements of the falloff in other galaxies. Though these measurements rely on emission from halo gas  kiloparsecs from the plane, the falloff parameter we measure is still remarkably consistent in magnitude. Additionally, in the Milky Way \citet{PLS2006} measured a falloff of 27 km s$^{-1}$ for a single structure 3.4 kpc above the plane; the falloff derived from this object is not directly comparable to our measurement. Since we measure the rolling motions and falloff parameter with $|z|\le$ 100 pc where the derivatives have magnitudes of  $\lesssim 10$ km s$^{-1}$, our result will also be difficult to confirm in other galaxies.  At higher $|z|$, \citet{FB2006} examine the halo gas kinematics in NGC 891 and NGC 2403 and conclude that the loss of angular momentum is due to the interaction between accreted low-angular momentum extragalactic gas and halo gas. It is conceivable that this effect is tied to the falloff we observe in the atomic disk gas, possibly with the vertical magnetic field playing a role. 
 Additionally, it may be important to consider rolling motions and the falloff parameter when examining gas kinematic phenomena such as supernova remnants and H {\footnotesize II} regions; the rotation curve derivatives can shear shells of gas if they are large enough.

\begin{deluxetable*}{ccccc}
\tablecaption{Falloff Parameter Measurements in Galactic Halos\label{table:summ}}
\tablehead{\colhead{Galaxy}&\colhead{Tracer} &\colhead{$d\Theta/d|z|$}&\colhead{Reference}}
\startdata
%Milky Way\tablenotemark{a}&H {\footnotesize I}, H $\alpha$&$-7.9$&\citet{PLS2006}\\
NGC 891&H {\footnotesize I}&$-25$ to $-100$&\citet{SSV1997}\\
NGC 891&H $\alpha$&$-15$ to $-18$&\citet{HRBB2006}\\
NGC 891&H {\footnotesize I}&$-15$&\citet{OFS2007}\\
NGC 891&H $\alpha$&$-18.8$&\citet{KPDV2007}\\
NGC 2403&H {\footnotesize I}&$-25$ to $-50$&\citet{FVSO2002}\\
NGC 4302&H $\alpha$, N {\footnotesize II}, S {\footnotesize II}&$-30$&\citet{HRBB2007}\\
NGC 5775&H $\alpha$&$-8$&\citet{HRBCB2006}
\enddata
\tablecomments{The derivative $d\Theta/d|z|$ is given in km s$^{-1}$ kpc$^{-1}$.} %\tablenotetext{a}{Based on a single feature.}}
\end{deluxetable*}

\section{Summary and Conclusions}

We fit an analytic form to the VGPS and SGPS H {\footnotesize I} spectra and  found the tangent velocity as a function of Galactic longitude and latitude. Using these tangent velocities, we fit a simple global model for the rotation curve and its first two vertical derivatives for lines of sight 100 pc or less from the plane. 
We then determined the rotation speed, rolling motion, and falloff in rotation speed from the plane as a function of Galactic  radius. We tested the accuracy of our fitting routines using a variety of simple H {\footnotesize I} model spectra. Our local parameter values are in agreement with our global fits. We demonstrated that the rolling motions are local effects, while the falloff from the plane has a consistent value of $-22\pm6$ km s$^{-1}$ kpc$^{-1}$, too large to be explained by gravitational physics alone, but consistent with the falloff measured in the halos of other galaxies.

\acknowledgements
Thanks to Naomi McClure-Griffiths for providing a copy of the SGPS, and to Bob Benjamin, Eliot Quataert, Eugene Chiang, Josh Peek, Kathryn Peek, Julia Comerford, Natalie Kittner, and Chris Hans for helpful conversations. E.S.L.~and L.B.~are supported by NSF grant AST-0540567. C.H.~is supported by NSF grant AST-0406987.

\bibliographystyle{apj}
%\bibliography{/Users/elevine/Documents/work/papers/mybiblio}

\end{document}